\documentclass[aps,prl,twocolumn, preprintnumbers,amsmath,amssymb,showpacs,superscriptaddress,amsfonts]{revtex4}
\usepackage{color}
\usepackage{graphicx}
\usepackage{amsmath}
\usepackage{amssymb}
\usepackage{epsfig}
\usepackage{wasysym}
\usepackage{bbm}
\usepackage{multirow}
\usepackage[colorlinks,citecolor=blue]{hyperref}
\renewcommand{\narrowtext}{\begin{multicols}{2} \global\columnwidth20.5pc}
\renewcommand{\v}[1]{{\bf #1}}

\newcommand{\s}{{\sigma}}

\def\be{\begin{eqnarray}}
\def\ee{\end{eqnarray}}
\newcommand{\Ref}[1]{Ref.~\cite{#1}}

\newcommand{\Eq}[1]{Eq.~(\ref{#1})}

\renewcommand{\>}{\rangle}

\newcommand{\e}{\epsilon}
\newcommand{\Fig}[1]{Fig.~\ref{#1}}

\newcommand{\imth}{\hspace{1pt}\mathrm{i}\hspace{1pt}}

\newcommand{\Rmnum}[1]{\expandafter\@slowromancap\romannumeral #1@}

\def\beq{\begin{equation}}
\def\eeq{\end{equation}}
\def\bpm{\begin{pmatrix}}
\def\epm{\end{pmatrix}}
\newcommand{\bal}{\begin{aligned}}
\newcommand{\eal}{\end{aligned}}

\newcommand{\bst}{{\boldsymbol{T}}}

\newcommand{\mbz}{{\mathbb{Z}}}
\newcommand \ti[1]{}

\begin{document}

\title{Can deeply underdoped superconducting cuprates be topological superconductors?}
\author{Yuan-Ming Lu}
\affiliation{Department of Physics, University of California,
Berkeley, CA 94720} \affiliation{Materials Sciences Division,
Lawrence Berkeley National Laboratory, Berkeley, CA 94720}
\author{Tao Xiang}
\affiliation{Institute of Physics, Chinese Academy of Sciences, Beijing 100190, China.}
\author{Dung-Hai Lee}
\affiliation{Department of Physics, University of California,
Berkeley, CA 94720} \affiliation{Materials Sciences Division,
Lawrence Berkeley National Laboratory, Berkeley, CA 94720}

%74.20.-z, 74.20.Rp, 71.27.+a}

\date{\today}

\begin{abstract}
The nodal $d_{x^2-y^2}$ superconducting gap is a hallmark of the cuprate high T$_c$ superconductors. Surprisingly recent angle-resolved photoemission spectroscopy of deeply underdoped cuprates revealed a nodeless energy gap which is adhered to the Fermi surface. Importantly this phenomenon is observed for compounds across several different cuprate families. In this letter we propose an exciting possibility, namely the fully gapped state is a topological superconductor.
\end{abstract}

\maketitle

The nodal $d_{x^2-y^2}$ gap function is a defining property of the copper-oxide (cuprate) high temperature superconductors. Many low temperature properties are affected because of the existence of gap nodes. According to topological arguments\cite{Volovik2003B,Horava2005,Wang2012,Matsuura2013} (see the supplementary material) the nodes should be stable against perturbations. Thus it is very surprising that recent angle-resolved photoemission spectroscopy (ARPES) experiments done on deeply underdoped cuprate samples (samples which are at the border between the antiferromagnetic (AF) phase and the superconducting (SC) phase)  have revealed a full particle-hole symmetric gap for Bi$_2$Sr$_2$CaCu$_2$O$_{8+\delta}$ (Bi2212)\cite{Tanaka2006,Vishik2012}, La$_{2-x}$Sr$_x$CuO$_4$ (LSCO)\cite{Ino2000,Razzoli2013}, Bi$_2$Sr$_{2-x}$La$_x$CuO$_{6+\delta}$ (Bi2201)\cite{Peng2013} and Ca$_{2-x}$Na$_x$CuO$_2$Cl$_2$ (NaCCOC)\cite{Shen2004}. The fact that there is a non-zero energy gap along the diagonal direction of the Brillouin zone (where the $d_{x^2-y^2}$ gap nodes usually sit) is referred to as the ``nodal gap'' phenomenon. Interestingly the nodal gap has been observed in systems whose magnetic and transport properties range from AF insulator\cite{Peng2013,Razzoli2013} to superconductor\cite{Vishik2012,Razzoli2013}. For Bi2212 \Ref{Vishik2012} proposes a phase diagram with a new superconducting (SC) phase appearing at the underdoping end of the SC dome. In contrast the samples showing the ``nodal gap'' in \Ref{Peng2013} are insulating and antiferromagnetic. However despite the difference in transport properties, relatively sharp coherence peaks (i.e. spectral peaks at the nodal gap edge) were observed in both systems so long as the doping concentration is not too low\cite{Vishik2012,Peng2013}. (Given the fact that samples at such a low doping level can be phase separated\cite{Fujita2012}, it is possible that the nodal gapped phase observed in \Ref{Peng2013}) lies in disconnect SC islands embedded in an insulating background.) In the literature possible cause of the nodal gap ranges from disorder induced Coulomb gap\cite{Chen2009} to the polaron effect\cite{Peng2013}.

Motivated by \Ref{Vishik2012,Peng2013,Razzoli2013} and the fact that non-superconducting gaps usually do not adhere to the Fermi surface, we {\it assert} that the state in question is a fully gapped SC state. Moreover we shall assume that it is not due to extrinsic effects such as disorder. Moreover, because samples exhibiting the nodal gap are exclusively found at the border between AF and SC we need to consider the possibility that such SC state coexists with the AF order. In the rest of the paper we first list all possible fully gapped SC states (with or without AF order), and organize them according to symmetry and topology (TABLE \ref{class}). This information will be combined with explicit effective theory calculations which determine the leading and subleading SC instabilities under different conditions. The combination of theses two approaches allows us to pin down the most likely candidate for the nodal gapped state, namely, a topological superconductor.

\begin{table*}[tb]
\centering
%\begin{ruledtabular}
\begin{tabular}{|c|c|c|c|c|c|}
  \hline
  % after \\: \hline or \cline{col1-col2} \cline{col3-col4} ...
  Neel order & Symmetry & Generators & Classification& AZ class& Examples \\ \hline
  No &$SU(2)_{spin}\times\bst$ & $\{\bst e^{\imth\pi S_x},\bst e^{\imth\pi S_y},\bst e^{\imth\pi S_z}\}$ & $\pi_0(R_5)=0$& CI& s-wave\\ \hline
  No &$SU(2)_{spin}$ & $\{e^{\imth\pi S_x}, e^{\imth\pi S_y}\}$ & $\pi_0(R_4)=\mbz$& C& $d\pm\imth d$\\ \hline
   No & $U(1)_{spin}$ &$\{e^{\imth\pi S_z}\}$ & $\pi_0(C_2)=\mbz$& A&$(p\pm\imth p)_{\uparrow\downarrow}$\\ \hline
   No & $\bst$ &$\bst$ & $\pi_0(R_1)=\mbz_2$& DIII& $(p\pm\imth p)_{\uparrow\uparrow}+(p\mp\imth p)_{\downarrow\downarrow}$\\ \hline
     No& None &N/A& $\pi_0(R_0)=\mbz$&D &$\alpha (p\pm \imth p)_{\uparrow\uparrow}+\beta (p\pm \eta \imth p)_{\downarrow\downarrow}$ \\ \hline
  Yes & $U(1)_{\rm spin}\times\bst e^{\imth\pi S_x}$&$\{\bst e^{\imth\pi S_x},\bst e^{\imth\pi S_y}\}$&$\pi_0(R_6)=0$&AI&$s$-wave\\ \hline
  Yes& $U(1)_{\rm spin}$ & $\{e^{\imth\pi S_z}\}$&$\pi_0(C_2)=\mbz$ &A &$(d\pm\imth d)$~;~$(p\pm\imth p)_{\uparrow\downarrow}$ \\ \hline
  Yes& None &N/A& $\pi_0(R_0)=\mbz$&D &$\alpha (p\pm\imth p)_{\uparrow\uparrow}+\beta (p\pm\eta\imth p)_{\downarrow\downarrow}$\\ \hline
\end{tabular}
\caption {Symmetry and topological classification of fully gapped SC phases in two spatial dimensions. We assume there is no spin-orbit interaction.  The 2nd  column lists the symmetry group whose generators are given in the 3rd column. The 4th column gives the Abelian group whose element each represents a topological class of SC states\cite{Kitaev2009,Wen2012} (for details see the supplementary material). $0$ means no topological superconductors, $\mbz_2$ means there exists one type of  topological superconductor in addition to the trivial (s-wave) superconductor. $\mbz$ represents the existence of infinite number of different topological superconductors each with protected gapless edge modes. The 5th column locates the symmetry class of each row in the Altland-Zirnbauer\cite{Altland1997} 10-fold way\cite{Schnyder2008,Kitaev2009}. The last column provides examples of gapped superconducting states in each symmetry class. Here $(p+\imth p)_{\uparrow\downarrow}$ denotes the $p+\imth p$ pairing between the spin up and spin down electrons, and $(p+\imth p)_{\uparrow\uparrow}, (p+\imth p)_{\downarrow\downarrow}$ represents $p+\imth p$ pairing among spin-up and and/or spin-down electrons. In row 6 and 9 $\alpha$ and $\beta$ denote generic complex numbers.}
\label{class}
%\end{ruledtabular}
\end{table*}

%Thus we start  by classifying possible fully gapped superconducting states with symmetry compatible with either paramagnetic or AF order in two dimensions.

Start with the maximally symmetric SC state, we systematically break down the SU(2)$_{\rm spin}\times\bst$ symmetry. (Because of superconductivity there is no U(1)$_{\rm charge}$ symmetry). For each residual symmetry we use the method of \Ref{Schnyder2008,Kitaev2009} to classify the possible fully gapped (two dimensional) SC states into topological classes. The result is TABLE \ref{class}, which contains eight cases in row 2 - 9. We group these cases according to whether superconductivity coexists with the AF order or not. (In view of the fact that the samples are likely disordered we do not consider crystal symmetries such as lattice translations. The only exception is that we do regard the system as having inversion symmetry, at least on average, so that spin singlet and spin triplet pairing will not mix. In the end we will briefly comment on the effects of inversion symmetry breaking.)

In the absence of Neel order there is spin SU(2) symmetry (in this paper we assume there is no spin-orbit interaction, which is a good approximation for the cuprates). In the singlet pairing case there are two classes of fully gapped SC state - the $s$-wave pairing (row 2) and $d\pm\imth d$ pairing (row 3). The latter is a topological SC state with chiral (complex) fermion edge modes. In fact \Ref{Razzoli2013} proposed $d\pm\imth d$ as an explanation of the nodal gap in LSCO.
In the triplet pairing case there are three classes of fully gapped SC state. They are listed in row 4-6 of TABLE \ref{class}. The $(p\pm ip)_{\uparrow\downarrow}$ SC state in the 4th row breaks the time-reversal symmetry but preserve U(1) spin rotation around, say, the z-axis. It is a representative of a family of degenerate triplet pairing state given by $\cos\theta (p\pm\imth p)_{\uparrow\downarrow}+\sin\theta e^{i\phi}(p\pm ip)_{\uparrow\uparrow}-\sin\theta e^{-i\phi}(p\pm ip)_{\downarrow\downarrow}$, where $\vec{d}=(\sin\theta\cos\phi,\sin\theta\sin\phi,\cos\theta)$ is the direction of the axis around which the U(1)$_{\rm spin}$ symmetry is preserved. These states span a manifold which is isomporphic to $S^2=\{(\sin\theta\cos\phi,\sin\theta\sin\phi,\cos\theta);\theta\in [0,\pi],\phi\in [0,2\pi)\}$.  The SC states in this class possess chiral (complex) fermion edge modes hence are topologically non-trivial.
In row 5 the $(p\pm ip)_{\uparrow\uparrow}+(p\mp ip)_{\downarrow\downarrow}$ SC state preserves the time-reversal symmetry but completely breaks the spin SU(2) symmetry. It has a pair of counter propagating Majorana modes along each edge. They are protected from back scattering by the time-reversal symmetry, hence $(p\pm ip)_{\uparrow\uparrow}+(p\mp ip)_{\downarrow\downarrow}$ is a time-reversal invariant topological superconductor.
The SC state in the 6th row of TABLE \ref{class} has no residual symmetry. In all but the $\eta=-1$ case there are  two chiral Majorana (equivalent to one complex) fermion edge modes. Hence they, too,  are chiral topological superconductors. If $\eta=-1$ the Majorana edge modes are counter-propagating. In this case an edge gap will open because with both time reversal and spin rotation symmetry broken nothing protects the counter-propagating Majorana edge modes from back scattering. Therefore  $\eta=-1$ represents the trivial element, i.e. 0, of $\mbz$.

The cases where the fully gapped SC state coexists with Neel order are listed in the last three rows of TABLE \ref{class}. Here, without loss of generality, we can assume the staggered magnetic moments to point in the $\pm z$-direction. The pairing states in the 7th row are all topologically trivial, they are exemplified by the $s$-wave pairing.  In contrast the  $d\pm\imth d$ and the $(p\pm\imth p)_{\uparrow\downarrow}$ SC states in the 8th row both give rise to chiral topological superconductors with chiral (complex) fermion edge modes. We note that the residual symmetry of the $(p\pm\imth p)_{\uparrow\downarrow}$ superconductor in row 8 is exactly the same as that in row 4. However, unlike row 4, there is no continuous degeneracy and associated Goldstone modes anymore because the SU(2)$_{\rm spin}$ is already broken down to U(1)$_{\rm spin}$ by the formation of the Neel order. %The superconductors in this row can be chiral and topological.
 The SC states in the 9th row are analogous to that given in the 6th row. Again, in all cases but $\eta=-1$  the state $\alpha (p\pm\imth p)_{\uparrow\uparrow}+\beta (p\pm\eta\imth p)_{\downarrow\downarrow}$ are topologically non-trivial.

Having surveyed all possible fully gapped SC states, the next step is to determine which class of TABLE \ref{class} does the experimentally observed fully gapped SC state belong to. To achieve that we use the effective theory of \Ref{Davis2013}
\be
H_{\rm eff}={\sum_{\v k}}'\sum_{\s}\e(\v k) \psi^+_{\s\v k}\psi_{\s\v k}+J\sum_{\<i,j\>}~\v S_i\cdot\v S_j.\label{eff}
\ee
It has been demonstrated that the \Eq{eff} is capable of capturing all experimentally observed electronic orders of the cuprates. Specifically the leading and subleading instabilities of \Eq{eff} in the particle-particle (Cooper pairing) and particle-hole (density wave and Fermi surface distortion) channels capture the d-wave superconductivity as well as the spin/charge density wave, nematicity and the $\v Q=0$ magnetic order\cite{Davis2013}.
Encouraged by such success we use it to predict the leading and subleading SC instabilities in the presence/absence of AF order. (See supplementary material for details.)

%In \Eq{eff} ${\sum_{\v k}}'$ stands for sum within a thin shell around the Fermi surface (so that $|\e(\v k)|$ is less than an energy cutoff). Moreover $\psi_{\s\v k}$ annihilates an electron with spin $\s$ in the band eigenstate with momentum $\v k$  in the thin shell. The last term in \Eq{eff} is written in real space but it should be understood as being projected to the low-energy band eigenstates in momentum space\cite{Davis2013}. In the following we shall work with this Hamiltonian and determine the leading and subleading superconducting instabilities. We will compare the results of the effective theory calculation with  TABLE \ref{class}.

\begin{figure}
\includegraphics[scale=.4]{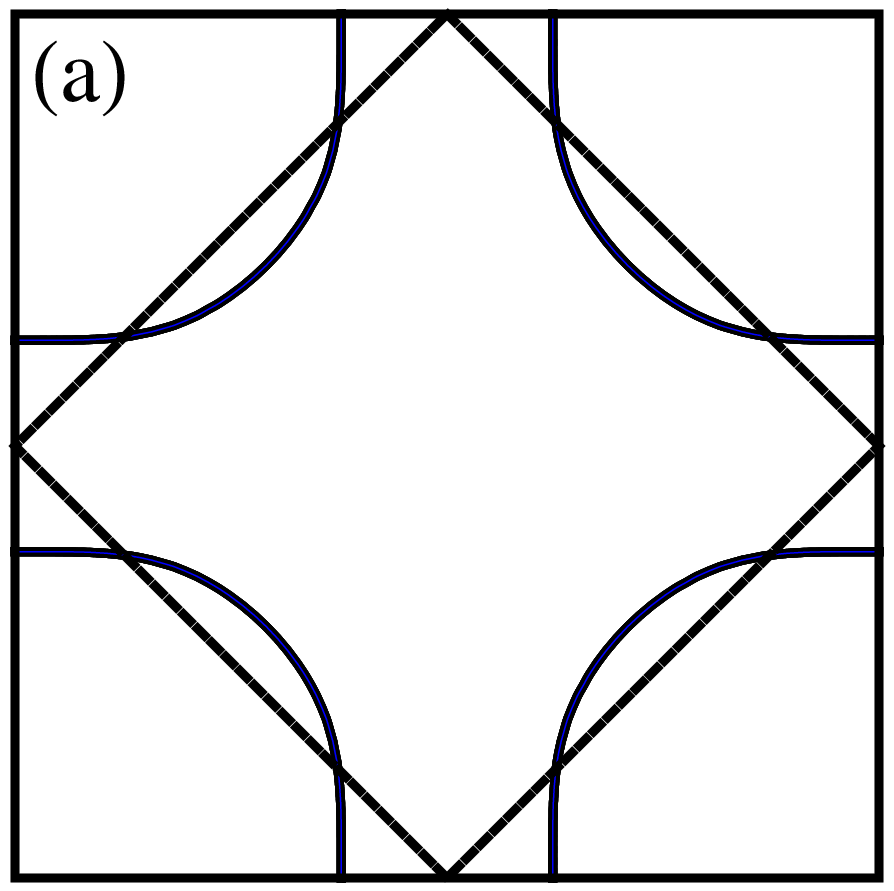}~~~\includegraphics[scale=.4]{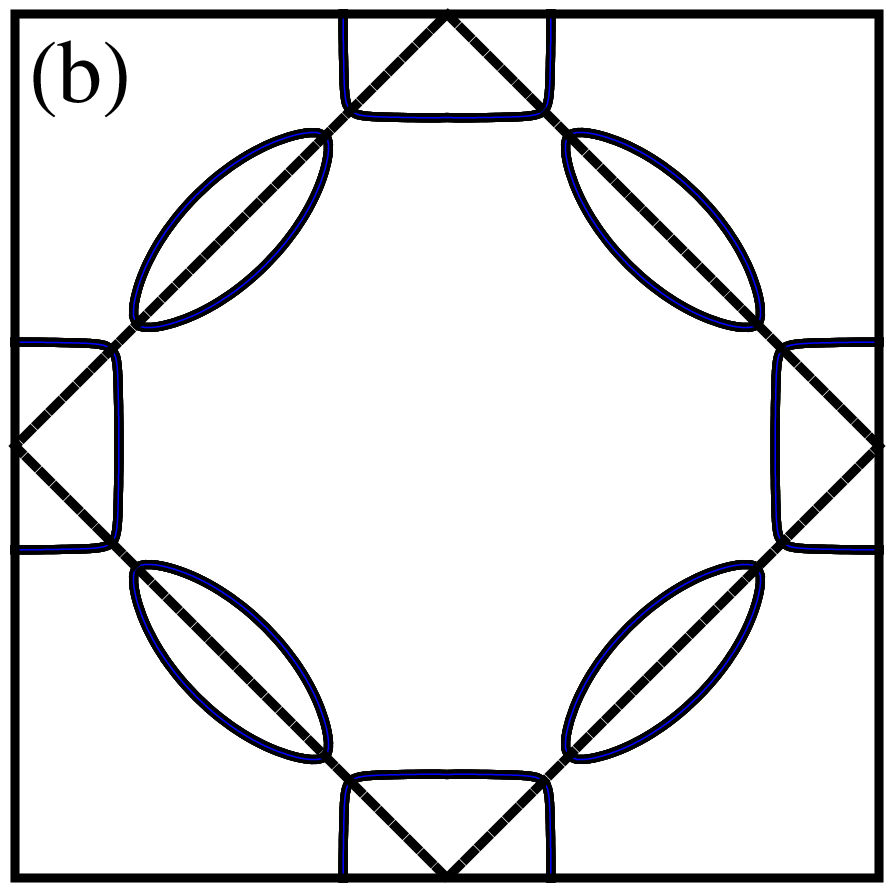}
\caption{The Fermi surface of cuprates without (a) and with (b) AF order. The dashed line enclose the AF Brillouin zone whoses vertices are $(\pm \pi/a,\pm \pi/a)$. The staggered moment used to construct panel (b) is $m=0.1$.}
\label{fs}
\end{figure}

The normal state Fermi surface with and without AF order are shown in \Fig{fs}.
In the AF phase the band dispersion and Bloch wavefunctions are the eigenvalues and eigenvectors of the following $2\times 2$ matrix
$
\left(
          \begin{array}{ll}
            \e_0(\v k) & \s~m \\
            \s~m & \e_0(\v k+\v Q) \\
          \end{array}
        \right).$
Here $\v Q=(\pi,\pi)$ is the Neel wavevector, $m$ is the staggered magnetization, and $\s=\pm 1$ labels the spin. $\e_0(\v k)$, the paramagnetic normal state dispersion, is given by $\mu-t_1(\cos k_x+\cos k_y)+t_2 \cos k_x\cos k_y-t_3 (\cos 2k_x+\cos 2k_y)$ with $t_1 = 1, t_2 = 0.3, t_3 = 0.2$ and $\mu = 0.14$. In the rest of the paper we shall set the value of $m$ to $0.1$. With this value of $m$ there are electron pockets centered around $(\pm\pi,0)$ and $(0,\pm\pi)$ and hole pockets centered around $(\pm\pi/2,\pm\pi/2)$.
 %It is worth to point out that although \Fig{fs}(b) is obtained using this specific value of $m$ the qualitative feature, namely the number and nature of the Fermi pockets and the location of their centers,  should not change as long as $m$ is not too big.

Based on \Eq{eff} and the bandstructure described above we decouple the AF interaction in all possible pairing channels and determine the gap functions that will first (and second) become unstable as $J$ is turned up from zero. (See the supplementary material.) %(The temperature is set at a value much lower than the energy cutoff of the momentum shell.
 Here a comment is in order. As mentioned earlier, the samples where the nodal gap is observed can be phase separated. However in our calculation translation symmetry is assumed. Therefore one should interpret our results as the local pairing instabilities. % which should apply locally to a disordered sample as well.
%leading and subleading superconducting pairing instabilities.
%The details are given in Ref.\cite{Davis2013}. In short, we decouple the AF interaction in all possible pairing channels and determine the gap functions that will first (and second) become unstable as $J$ is turned up from zero.  %Under this condition the $U(1)$ rotation (which is denoted as $U(1)_{S_z}$ in TABLE \ref{class}) around the z-axis in spin space is a good symmetry. Here the hatch size is proportional to the magnitude of the gap and the color (red: negative, blue: positive) denotes the sign.

\begin{figure}
\includegraphics[scale=.3]{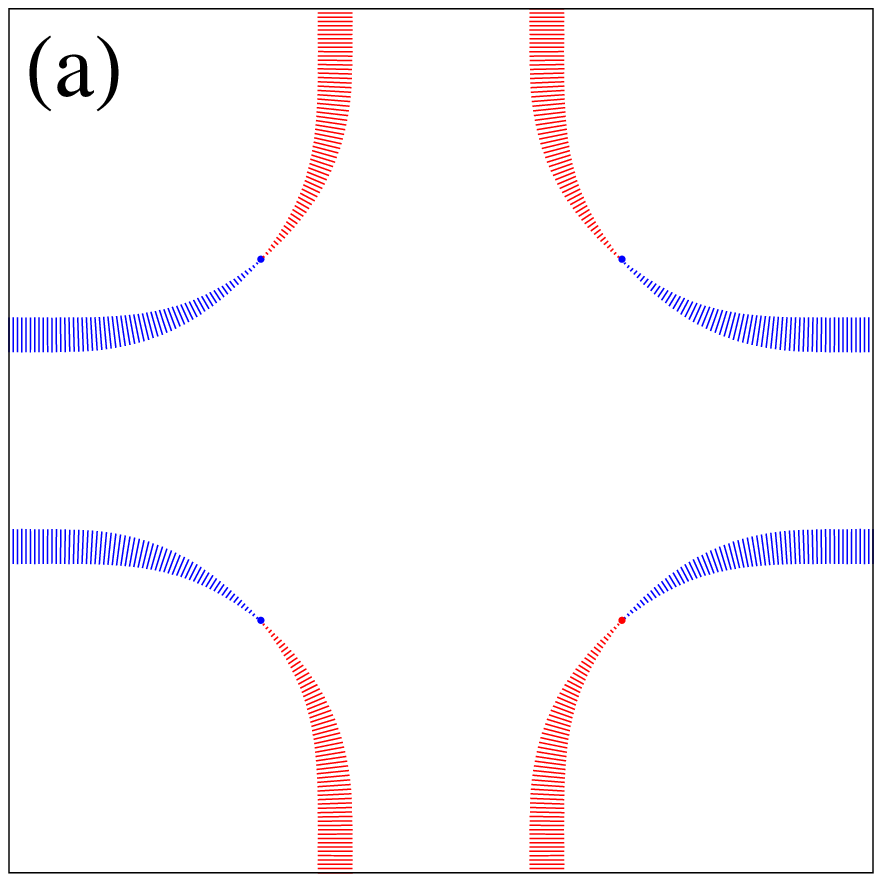}\includegraphics[scale=.3]{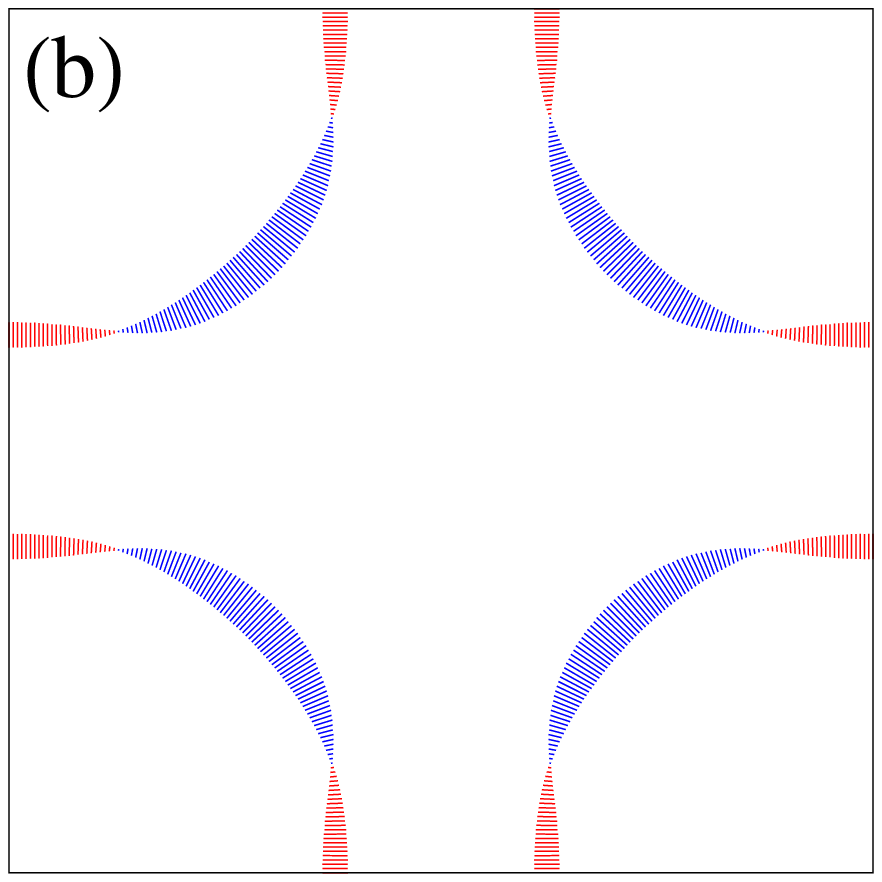}\includegraphics[scale=.3]{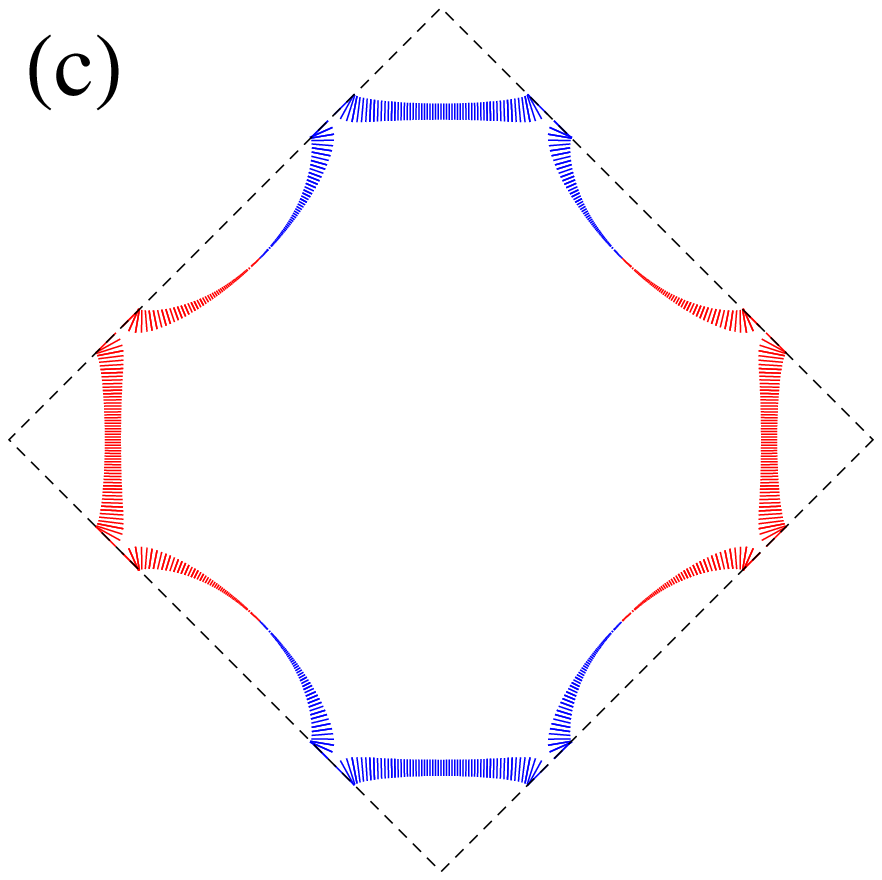}
\includegraphics[scale=.3]{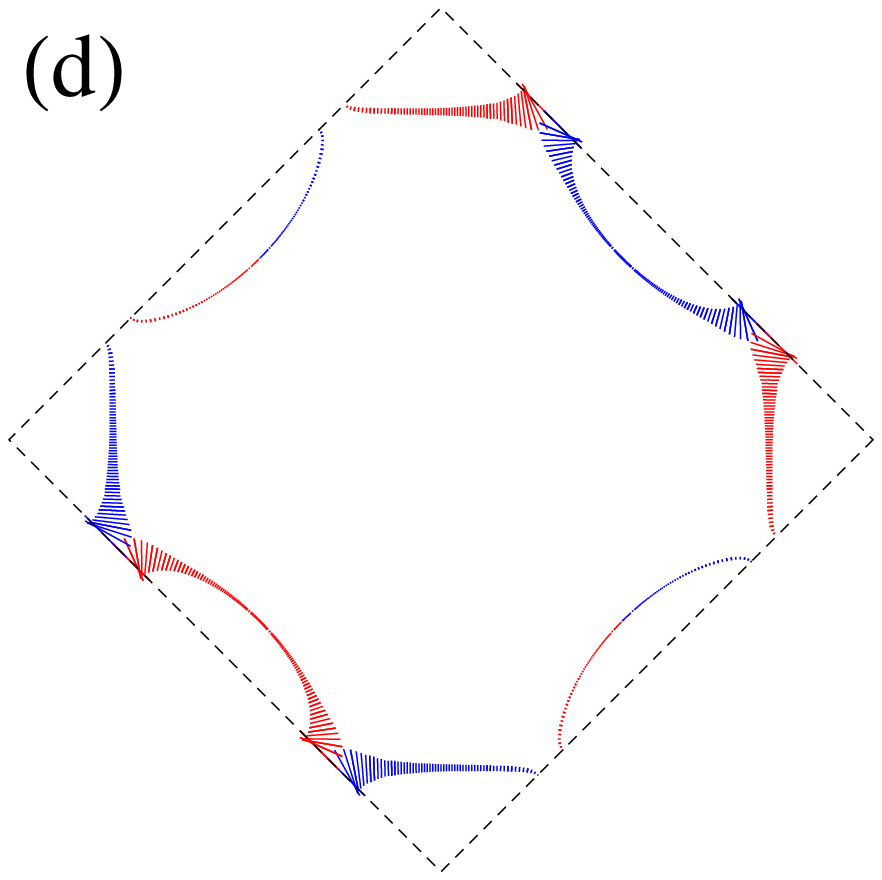}\includegraphics[scale=.3]{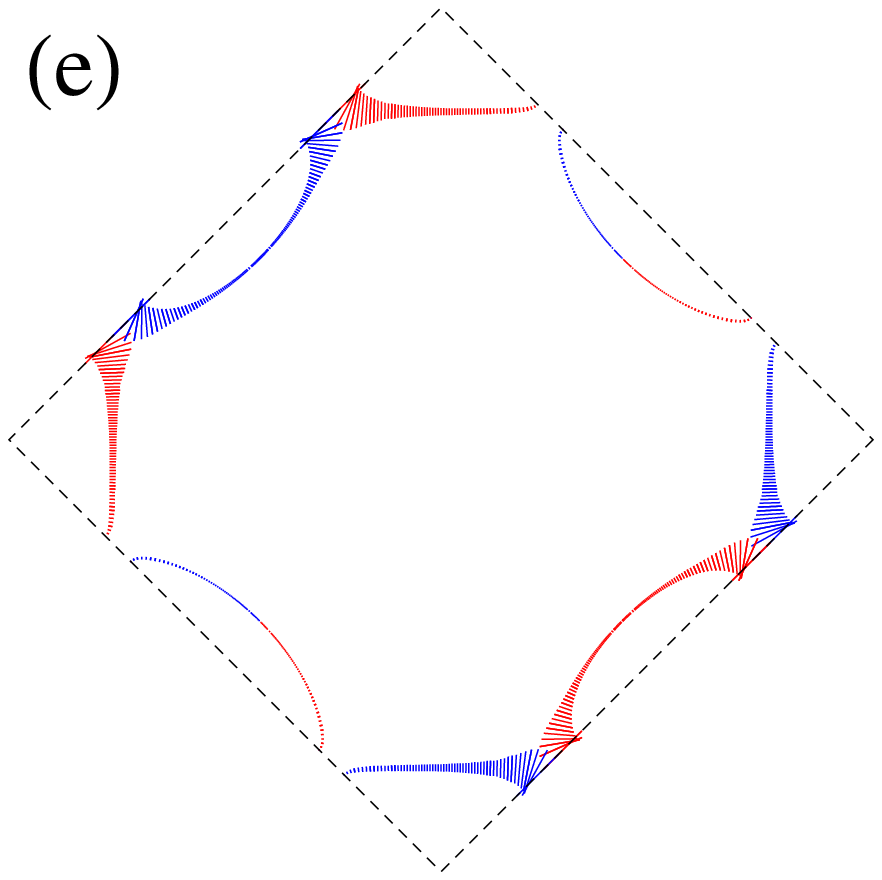}\includegraphics[scale=.3]{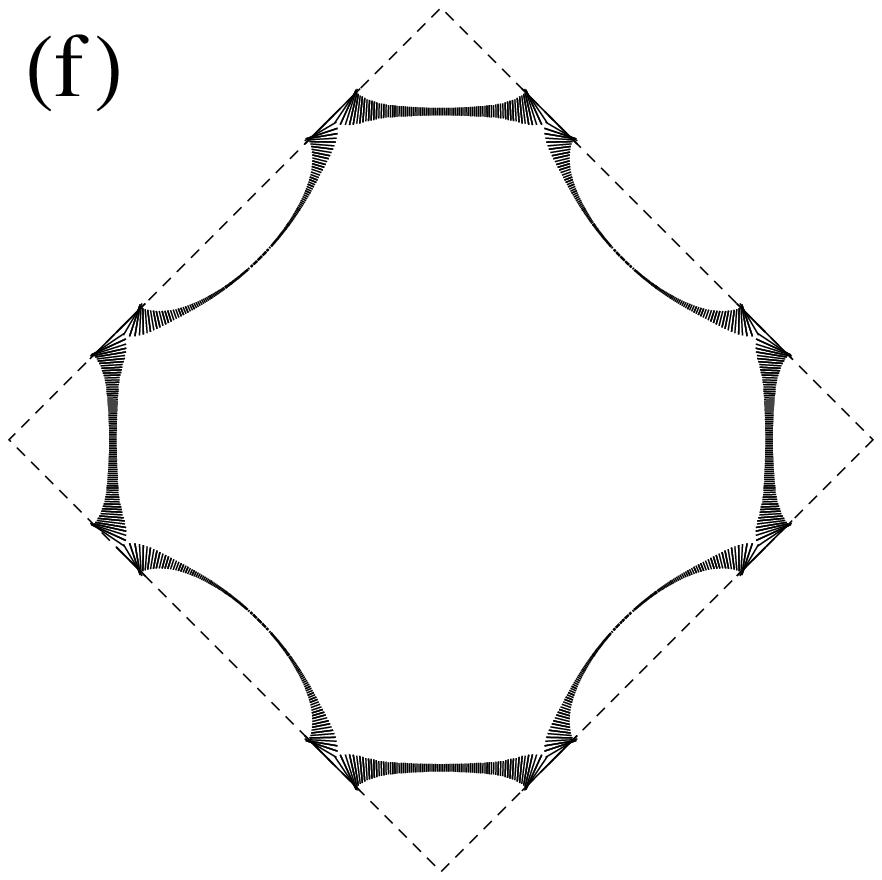}
\caption{The leading superconducting instabilities of \Eq{eff} in the absence (a)-(b) and presence (d)-(f) of Neel order. (a) singlet $d_{x^2-y^2}$ symmetry, (b) extended $s$ symmetry,  (c) singlet $d_{x^2-y^2}$ symmetry, (d) $p_{x+y}$ and (e) $p_{x-y}$ symmetries. Here the hatch size is proportional to the magnitude of the gap and the color indicates the sign (red: negative, blue: positive). (f) The energy gap correspond to $|\Delta_d(\v k)+\imth \Delta_e(\v k)|$. Here Black means positive and the hatch size is proportional to the gap magnitude.}
\label{gap}
\end{figure}

{\bf{%First we discuss the case of no AF order and the
(i) Cooper pairing in the absence of AF order.}}  %(where the two electrons in the Cooper pair have opposite spins).
\Fig{gap}(a) and (b) illustrate the leading and subleading SC instabilities in the absence of AF order\cite{Davis2013}. The $d_{x^2-y^2}$ pairing in panel (a) has four nodes, hence can not be responsible for the fully gapped state observed in experiments. Panel (b) illustrates the subleading extended s-wave pairing instability. It, too, has nodes. These nodes, like those of the $d_{x^2-y^2}$ pairing are topologically stable against perturbations. Although they are not required by the point group symmetry, it requires, e.g. strong disorder, to get rid of them (the same for the $d_{x^2-y^2}$ nodes). Given the fact that the $d_{x^2-y^2}$ nodes have been observed in very disordered samples, we regard it unlikely that the extended s-wave pairing instability is responsible for the nodal gap. In the triplet pairing channel \Eq{eff} has no SC instability (as long as $J>0$). Thus row 2-6 of TABLE \ref{class} are ruled out on the basis that at least one component (e.g. the $d_{xy}$ of $d_{x^2-y^2}+\imth d_{xy}$) of the gap function is not among the top (i.e. leading or subleading) pairing instabilities in \Fig{gap}(a,b).

{{\bf (ii) Cooper pairing in the presence of AF order.}} In this case the leading and subleading pairing instabilities occurs in the $S_z=0$ channel and are shown in \Fig{gap}(c)-(e). The leading paring symmetry is again $d_{x^2-y^2}$, which can not account for the presence of nodal gap. The subleading pairing symmetries, $p_{x+y}$ and $p_{x-y}$ in \Fig{gap}(d,e), are degenerate. However although they each has nodes, the linear combination  $(p\pm \imth p)_{\uparrow\downarrow}$  can give rise to a fully gapped chiral topological superconductor. This superconductor belong to the topological class of the 8th row of TABLE \ref{class}  Therefore combining TABLE \ref{class} with explicit calculations we conclude that the best candidates for the experimentally observed fully gapped state is the $(p\pm\imth p)_{\uparrow\downarrow}$ SC coexists with AF order (the 8th row of TABLE \ref{class}).

 According to \Ref{Vishik2012,Peng2013,Razzoli2013} the nodal gap magnitudes increases as $\v k$ moves away from the diagonal direction. This is qualitatively consistent with the behavior of $|\Delta_d(\v k)+\imth \Delta_e(\v k)|$  (see \Fig{gap}(f)) where $\Delta_{d,e}(\v k)$ are the gap functions of \Fig{gap}(d).
\begin{figure}
\includegraphics[scale=.8]{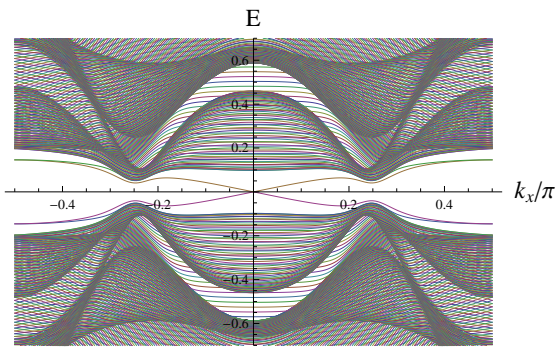}
\caption{The edge spectrum of the  $(p+\imth p)_{\uparrow\downarrow}$-AF coexisting state. Here due to the Brillouin zone folding $k_x=-\pi/2$ and $k_x=\pi/2$ are equivalent.The left/right moving branches are localized on opposite edges. This figure is constructed by fitting the numerical gap functions in \Fig{gap}(d,e) to simple trigonometric functions, then use the fitted function in the real space edge calculation.}
\label{edge}
\end{figure}
 In \Fig{edge} we show the edge spectrum of the SC state discussed above. Explicit wavefunction calculation shows the left/right moving in-gap modes are localized on opposite edges. However despite the presence of edge states, we do not expect the superconducting vortex to harbor zero modes. This is because in one dimension (the dimension of a loop surrounding the vortex) the symmetry class of row 8 of TABLE \ref{class} only has trivial states (see supplementary materials).

{\bf Discussions:} A natural question one might ask is why does triplet pairing instability exist in the AF state but not in the paramagnetic state? It turns out that this is related to the unit cell doubling in the AF state. After such doubling $\v Q=(\pi,\pi)$ becomes a reciprocal lattice vector. Hence center-of-mass (COM) momentum $(0,0)$ and $(\pi,\pi)$ Cooper pairing can coexist. In the AF state although the AF exchange interaction in \Eq{eff} favors singlet pairing for COM momentum $(0,0)$, it favors triplet pairing if the COM momentum is $(\pi,\pi)$. %The fact that we have either singlet or triplet pairing (not linear combination of them) in \Fig{gap}(c-e) is due to the inversion symmetry.
One might also ask ``under what condition will the triplet pairing in \Fig{gap}(d,e) become the leading instability?'' It turns out that this can be achieved by
increasing the staggered moment $m$, or by slight modifying the bandstructure so that in the AF state there is only hole pockets. For example by using $t_1 = 1, t_2 = 0.3, t_3 = 0, \mu = 0.2$ and $m=0.1$ the triplet pairing can be made to surpass the singlet one when a nearest neighbor repulsion $V$ with $V/J\gtrsim 1.4$ is added to \Eq{eff}. %This shows that it is possible that the triplet pairing is stabilized.
Given the fact that AF order induces triplet pairing instability, an interesting question arises. Is it possible that  strong fluctuations of the Neel order parameter (which should exist near the AF-SC boundary) can stabilize the triplet topological SC order even when there is no static AF order? Another interesting issue is the destruction of SC coherence due to the orientation fluctuations of the AF order parameter. We have seen that in the presence of Neel order the infinite pairing degeneracy ($\vec{d}\in S^2$) in the 4th row of TABLE \ref{class} is reduced to the two fold degeneracy ($\vec{d}\parallel\pm \hat{z}$) in the 8th row of TABLE \ref{class}.  (In fact the overall sign of $\vec{d}$ can be absorbed by the charge U(1) phase of superconductivity, hence there is really no degeneracy left). The preceding fact implies the direction of the Neel order parameter pins the $\vec{d}$ vector of triplet pairing. If so it is reasonable to expect the orientation fluctuations of the Neel order parameter can impede (or even destroy) the SC coherence. This reasoning suggests that strong fluctuation of the Neel order can stretch the regime of fluctuation SC (hence the ``pseudogap region'') to a much wider temperature interval. This might be related to the observation of pseudogap above the SC transition\cite{Vishik2012}. Lastly we comment on the effects of inversion symmetry breaking. Without inversion symmetry singlet (\Fig{gap}(c)) and triplet (\Fig{gap}(d,e)) pairing channels can mix. This can occur, e.g., locally due to disorder or phase separation. We have checked that the superconductor with $\alpha d_{x^2-y^2}+\beta (p\pm\imth p)_{\uparrow\downarrow}$ can be fully gapped. In addition there is a phase transition from a topologically trivial to non-trivial phase as $|\beta/\alpha|$ increases. The details of this investigation will be published in future publication.

{\bf Conclusion:} We propose that deeply underdoped cuprates might be a topological superconductor. One way to experimentally test our prediction is to use STM to image the edge states. Given the likelihood that the sample is phase separated, STM is a particularly valuable probe for the signature of the topological superconductivity locally.

\begin{acknowledgments}
We thank Yu He and Makoto Hashimoto for useful discussions. This work is supported by DOE Office
of Basic Energy Sciences, Division of Materials Science, grant
DE-AC02-05CH11231 (YML,DHL).
\end{acknowledgments}

%\bibliographystyle{apsrev}
%\bibliography{F://Research//Ctex//Bib//mybib//bibs}
%\bibliography{E://Dropbox//notes//bibs}

%\begin{references}
%\bibitem{volovik} G.E. Volovik ``The universe in a helium droplet'',  Oxford University Press (2003).
%\bibitem{horava} P. Ho$\check{r}$ava, Phys. Rev. Lett. {\bf95}, 016405 (2005).
%\bibitem{Tanaka}K. Tanaka {\it et al.}, Science, {\bf314}, 1910 (2006).
%\bibitem{vishik} I.M. Vishik {\it et al}, PNAS {\bf 109}, 18332-18337 (2012).
%\bibitem{Ino} A. Ino {\it et al}, Phys. rev. B {\bf62}, 4137(2000).
%\bibitem{swiss} E. Razzoli {\it et al}, Phys. Rev. Lett. {\bf 110}, 047004 (2013).
%\bibitem{xj} Y. Peng {\it et al}, Nature commun. {\bf 4}, 2459 (2013).
%\bibitem{kyle} K.M. Shen {\it et al}, Phys. Rev. {\bf 69}, 054503 (2003).
%\bibitem{disorder} K. Bouadim {\it et al}, Nature Physics {\bf 7}, 884 (2011).
%\bibitem{separation} For an experimental review see, e.g.,  K. Fujita {\it et al}, J. Phys. Soc. Jpn. {\bf81}, 011005 (2012).
%\bibitem{kitaev}
%A. Y. Kitaev,
%\jn{AIP Conf. Proc.} {\bf 1134}, 22 (2009).
%\bibitem{ryu}
%A. P. Schnyder, S. Ryu, A. Furusaki, and A. W. W. Ludwig,
%\jn{Phys. Rev. B} {\bf 78}, 195125 (2008);
%S. Ryu, A. P. Schnyder, A. Furusaki, and A. W. W. Ludwig,
%\jn{New J. Phys.} {\bf 12}, 065010 (2010).
%\bibitem{kivelson} E. Fradkin nd S.A. Kivelson SA, Nature Physics, {\bf8}, 864-866 (2012).
%\bibitem{davisLee} J.C. Davis and D.-H. Lee, PNAS
%
%\end{references}

\end{document}

\title{Supplementary materials}
\author{Yuan-Ming Lu}
\affiliation{Department of Physics, University of California,
Berkeley, CA 94720} \affiliation{Materials Sciences Division,
Lawrence Berkeley National Laboratory, Berkeley, CA 94720}
\author{Tao Xiang}
\affiliation{Institute of Physics, Chinese Academy of Sciences, Beijing 100190, China.}
\author{Dung-Hai Lee}
\affiliation{Department of Physics, University of California,
Berkeley, CA 94720} \affiliation{Materials Sciences Division,
Lawrence Berkeley National Laboratory, Berkeley, CA 94720}

\date{\today}

%%%%%%%%%%%%%%%%%%%%%%%%%%%%%%%%%%%
%\begin{abstract}
%\end{abstract}
%
%
%\pacs{}

\maketitle

\section{Stability of nodal quasiparticles}

In a singlet superconductor with no magnetic orders, the system preserves $SU(2)$ spin rotational symmetry and time reversal symmetry $\bst$. A generic nodal particle in this 2+1-D time-reversal-invariant (TRI) singlet superconductor can be described as
\bea
H_{{\bf Q}+{\bf k}}=k_x\gamma_1+k_y\gamma_2.
\eea
where we have normalized the Fermi velocities to unity without loss of generality. ${\bf k}$ is a small momentum expanded around the nodal point ${\bf Q}$. Usually the nodal Hamiltonian $H_{{\bf Q}+{\bf k}}$ is written in complex fermion basis $\Psi=(c,c^\dagger)^T$, but we can also represent it in the Majorana basis\cite{Kitaev2009} $\eta$ through relation $c=\eta_1+\imth\eta_2$. In the Majorana fermion basis $\gamma_{1,2}$ are real symmetric matrices which square to be $+1$\cite{Wen2012}. Time reversal symmetry for spin=$1/2$ electrons looks like
\bea
{\bf T}\eta_i{\bf T}^{-1}=T_{i,j}\eta_j,~~~T^2=-1.
\eea
where $T$ is again a real matrix. Meanwhile spin rotational symmetries are
\bea
e^{\imth\pi S_\alpha}\eta_ie^{-\imth\pi S_\alpha}=\big(\Sigma_\alpha\big)_{i,j}\eta_j,~~~(\Sigma_\alpha)^2=-1.
\eea
where $S_\alpha,~\alpha=x,y,z$ are the total spin operators. Note that $[T,\Sigma_\alpha]=0$. A symmetric nodal Hamiltonian $H_{{\bf Q}+{\bf k}}$ obey
\bea
\{\gamma_i,T\}=[\gamma_i,\Sigma_\alpha]=0.
\eea
The symmetry group $SU(2)_s\times\bst$ can be generated by the following 3 generators
\bea\label{nodal symmetry}
\bst e^{\imth\pi S_\alpha}\rightarrow T_\alpha\equiv T\Sigma_\alpha,~~~(T_\alpha)^2=1,~~~\{T_\alpha,\gamma_i\}=0.
\eea
Notice that matrices $\{\gamma_i,T_\alpha\}$ form Clifford algebra
The problem of classifying different nodal Hamiltonians $H_{{\bf Q}+{\bf k}}$ reduces to the following mathematical question: what are the distinct classes of real matrix $\gamma_{1}$ (with $\gamma_2$ and $T_\alpha$ fixed) satisfying condition (\ref{nodal symmetry})? This question is related to (real) Clifford algebra $Cl_{5,0}$ and the associated classifying space for $\gamma_1$ is $R_4=\big(Sp(k+m)/Sp(k)\times Sp(m)\big)\times\mbz$ in notation of \Ref{Kitaev2009}. Therefore the symmetry classification of nodal Hamiltonian $H_{{\bf Q}+{\bf k}}$ with symmetry (\ref{nodal symmetry}) is given by $\pi_0(R_4)=\mbz$. An intuitive way to understand this $\mbz$ classification is the following\cite{Matsuura2013}: each node also appears on the surface of a 3+1-D TRI singlet superconductor. As a result the symmetry classification of distinct 2+1-D nodal Hamiltonians is the same as symmetry classification of 3+1-D gapped TRI singlet superconductors (class CI\cite{Schnyder2008,Kitaev2009}) with protected surface states.

Now we show how to compute this integer ``winding number" for a given nodal Hamiltonian. In the presence of $SU(2)_s\times\bst$ symmetry, the quadratic electron Hamiltonian near the node at ${\bf Q}$ can always be written as
\bea
&\notag H_{\bf Q}=\sum_{\bf k}\bpm c^\dagger_{{\bf Q+k},\uparrow}\\c_{-{\bf Q+k},\downarrow}\epm^T h_{\bf k}\bpm c_{{\bf Q+k},\uparrow}\\c^\dagger_{-{\bf Q+k},\downarrow}\epm\\
&+\bpm c^\dagger_{{\bf Q+k},\downarrow}\\c_{-{\bf Q+k},\uparrow}\epm^T \tau_yh_{\bf k}\tau_y\bpm c_{{\bf Q+k},\downarrow}\\c^\dagger_{-{\bf Q+k},\uparrow}\epm
\eea
where $\tau_{x,y,z}$ represent Pauli matrices in the particle-hole channel. Let's label the filled bands (negative energy eigenstate) of Hermitian matrix $h_{\bf k}$ by $|{\bf k},\alpha\rangle$, then the integer winding number $w$ for nodal point ${\bf Q}$ is given by
\bea
w=\frac\imth{2\pi}\oint_{\bf Q}\text{d}{\bf k}\text{Tr}\big[\partial_{\bf k}\hat{P}_{\bf k}\big],~~~\hat{P}_{\bf k}\equiv\sum_{\alpha~\text{filled}}|{\bf k},\alpha\rangle\langle{\bf k},\alpha|.
\eea
It's straightforward to check\cite{Wang2012} that $w=\pm2$ for each node in a $d_{x^2-y^2}$-wave cuprate superconductor.

\section{Classification of gapped magnetic superconductors}

In this section we derive the symmetry classification presented in TABLE I of the main text. Following \Ref{Kitaev2009}, as far as topological properties are concerned, a gapped phase in two spatial dimensions can always be obtained by adding a mass matrix $M$ to the Dirac Hamiltonian
\bea
H_0=\imth\partial_x\gamma_1+\imth\partial_y\gamma_2.
\eea
In the Majorana basis $\gamma_{1,2}$ are real symmetric matrices which square to be $+1$. The mass terms correspond to real antisymmetric matrices $M$ satisfying $\{M,\gamma_i\}=0$. If the mass term $M$, matrices $\gamma_{1,2}$ and all symmetry group generators form real clifford algebra $Cl_{p,1}$, the associated classifying space\cite{Kitaev2009,Wen2012} for mass $M$ is $R_{2-p\mod8}$.

In the presence of $SU(2)$ spin rotational symmetry and time reversal symmetry $\bst$, the symmetry group for non-interacting fermions is generated by time reversal followed by $\pi$-spin-rotations along the 3 axes: $T_\alpha=\bst e^{\imth\pi S_\alpha}$. In the Majorana basis the $T_\alpha$ are real symmetric matrices satisfying $\{T_\alpha,T_\beta\}=2\delta_{\alpha,\beta}$ and $\{T_\alpha,\gamma_i\}=0$. Therefore the symmetry-allowed $M$ form the real Clifford algebra $Cl_{5,1}$ with $\gamma_i$ and $T_\alpha$ and hence belong to classifying space $R_5$. Since $\pi_0(R_5)=0$ there is only trivial mass, hence there are no topological superconductors with $SU(2)\times\bst$ symmetry in two space dimensions. This corresponds to Altland-Zirnbauer class\cite{Zirnbauer1996,Altland1997} CI in the 10-fold way of topological insulators/superconductors\cite{Schnyder2008}.

When time reversal is spontaneously broken but spin $SU(2)$ symmetry is preserved, the generators of symmetry are $\{e^{\imth\pi S_x},e^{\imth\pi S_z}\}$. In the Majorana basis the two generators are again real anti-symmetric matrices $\Sigma_{x,z}$ both of which square to $-1$. Without loss of generality, we can choose the two generators to be $\Sigma_{x,z}=\sigma_{x,z}\otimes(\imth\sigma_y)\otimes1_{N\times N}$ ($\vec\sigma$ are the Pauli matrices and $N$ is an arbitrary integer), and the mass matrix $M$ must have the following form
\bea
&\notag M=\sigma_0\otimes\sigma_0\otimes H_0+\imth\sigma_y\otimes\sigma_x\otimes H_x+\sigma_0\otimes(\imth\sigma_y)\otimes H_y+\imth\sigma_y\otimes\sigma_z\otimes H_z,\\
&\notag H_0^T=-H_0,~~~H_\alpha^T=H_\alpha~(\alpha=x,y,z).
\eea
so that $[\Sigma_{x,z},M]=0$. It's straightforward to show that $M^2=-1$ implies
\bea
(H_0)^2-\sum_{\alpha=x,y,z}(H_\alpha)^2=-1_{N\times N},~~~\{H_0,H_\alpha\}=0,~~~[H_\alpha,H_\beta]=0.
\eea

Such a mass term $M$ can be mapped into a quaternion matrix\cite{Wen2012}
\bea
H=H_0+\imth H_x+\jmth H_y+\kmth H_z,~~~~~H^2=-1_{N\times N}.
\eea
where $\imth,\jmth,\kmth$ satisfy the quaternion algebra. Meanwhile notice that the presence of 4 anti-commuting symmetry generators (squaring to be $+1$) also lead to the classifying space of a quaternion matrix
\bea
&\notag\Gamma_{1,2}=\sigma_{x,z}\otimes\sigma_0\otimes\sigma_0\otimes1_{N\times N},~~~\Gamma_{3,4}=\sigma_y\otimes\sigma_y\otimes\sigma_{x,z}\otimes1_{N\times N},~~~\{M,\Gamma_i\}=0\Longrightarrow\\
&M=1\otimes H_0+\imth\otimes H_x+\jmth\otimes H_y+\kmth\otimes H_z,\\
&\notag1\equiv\imth\sigma_y\otimes\sigma_y\otimes\sigma_y,~\imth\equiv\imth\sigma_y\otimes\sigma_x\otimes\sigma_0,
~\jmth\equiv\imth\sigma_y\otimes\sigma_z\otimes\sigma_0,
~\kmth\equiv\sigma_y\otimes\sigma_0\otimes\sigma_y.
\eea
After taking the two Dirac matrices $\gamma_{1,2}$ into account ($\gamma_{1,2}^2=1$), we can see all matrices $\{M,\gamma_{1,2},\Gamma_i\}$ form Clifford algebra $Cl_{6,1}$, which leads to classifying space $R_4=\frac{Sp(l+m)}{Sp(l)\times Sp(m)}\times\mbz$ for $SU(2)_{spin}$-symmetric superconductors in 2+1-D. Since $\pi_0(R_4)=\mbz$, distinct singlet topological superconductors are labeled by an integer \ie their spin quantum Hall conductance\cite{Senthil1999}. This corresponds to AZ class C.

When spin rotational symmetry is spontaneously broken while preserving time reversal symmetry ($\bst^2=-1$), there is only one generator $T$ satisfying $T^2=-1$. Therefore the matrices $\{\gamma_{1,2},T,M\}$ form Clifford algebra $Cl_{2,2}$, which leads to classifying space\cite{Kitaev2009,Wen2012} $R_1=O(n)$. Since $\pi_0(R_1)=\mbz_2$, there is only one type of topologically nontrivial superconductor with time reversal symmetry ($\bst^2=-1$) in addition to the topologically trivial ones. This corresponds to AZ class DIII.

When AF Neel order (along z-axis) coexists with superconductivity, the symmetry group reduces to $U(1)$ spin rotation along z-axis and time reversal combined with an in-plane $\pi$-spin-rotation. This symmetry group for non-interacting fermions is generated by $\bst e^{\imth\pi S_x}$ and $\bst e^{\imth\pi S_y}$ (since $e^{\imth\pi S_z}=\bst e^{\imth\pi S_x}\bst e^{\imth\pi S_y}$, we don't need to write down $e^{\imth\pi S_z}$ additionally). Again in Majorana basis these two generators are real symmetric matrices $T_\alpha$ with $\{T_\alpha.T_\beta\}=2\delta_{\alpha,\beta}$. Therefore the real Cliford algebra formed by mass $M$ and $\{\gamma_{1,2},T_{x,y}\}$ is $Cl_{4,1}$ and the symmetry-allowed mass $M$ belong to classifying space $R_6$. Since $\pi_0(R_6)=0$ there are no topologically nontrivial superconductors protected by this symmetry. It corresponds to AZ class AI.

When the symmetry further breaks down to just $U(1)$ spin rotation along the axis of collinear AF Neel order, say the $z$-axis, the symmetry group is generated by $\pi$-spin-rotation along $z$-axis. In the Nambu basis $\psi=(c_\uparrow,c^\dagger_\downarrow)$ the quadratic gapped Hamiltonian has the following form:
\bea
&\notag H=\imth\partial_x\gamma_1+\imth\partial_y\gamma_2+M,\\
&\gamma_i^2=1,~~~M^2=+1,~~~\{M,\gamma_i\}=0.
\eea
where we've normalized the fermi velocities and the size of mass term $M$. The classifying space for $M$ is $C_2=\frac{U(m+n)}{U(m)\times U(n)}\times\mbz$ and we have $\pi_0(C_2)=\mbz$. This means there is an infinite number of distinct topological superconductors with $S_z$ conservation in 2+1-D and they are labeled by an integer. This integer is nothing but the Chern number\cite{Thouless1982} of quadratic Hamiltonian $H$, associated to the number of chiral (complex) fermion modes on the edge. This corresponds to AZ class A.

When this $U(1)$ spin rotation is further broken, there is no symmetry in the system. The mass term $M$ together with Dirac matrices $\gamma_{1,2}$ forms clifford algebra $Cl_{2,1}$, and the classifying space for mass term $M$ is $R_0=\frac{O(m+n)}{O(m)\times O(n)}\times\mbz$. Since $\pi_0(R_0)=\mbz$ there are an infinite number of topological superconductors in two space dimensions with no symmetry. They are labeled by an integer, which equals the number of chiral Majorana mode on the edge. This corresponds to AZ pclass D.

In the end we comment on vortex bound states in these superconductors. Notice that a superconducting vortex always breaks time reversal symmetry, Meanwhile the existence of protected in-gap bound state at a point defect (the vortex core) is determined by the symmetry classification in one space dimension less\cite{Teo2010,Wen2012}. Since the topological classification is always trivial for class A in one spatial dimension, the topological $(p+\imth p)_{\uparrow,\downarrow}$ superconductors with chiral edge modes don't support zero-energy vortex bound state.

\section{Effective theory calculations}

We follow the procedures of \Ref{Davis2013} to determine the top SC instabilities from the effective Hamiltonian. Starting with
\be
H_{\rm eff}={\sum_{\v k}}'\sum_{\s}\e(\v k) \psi^+_{\v k\s}\psi_{\v k\s}+J\sum_{\<i,j\>}~\v S_i\cdot\v S_j.\label{S1}
\ee
In \Eq{S1} ${\sum_{\v k}}'$ stands for sum within a thin shell around the Fermi surface (so that $|\e(\v k)|$ is less than an energy cutoff). Moreover $\psi_{\s\v k}$ annihilates an electron with spin $\s$ in band eigenstate at momentum $\v k$  within the momentum shell. Note in $\psi_{\s\v k}$ we did not keep the band indices. This is because with the restriction to a thin energy shell, momentum actually fixes the band index.

First we rewrite the last term of \Eq{S1} in momentum space. In view of the fact that we also need to treat pairing in the AF ordered state, we use the folded, i.e. the magnetic, Brillouin zone (MBZ). (We use MBZ even when treating the paramagnetic state).
\be &&V_{\rm int}=J\sum_{\<i,j\>}~\v S_i\cdot\v S_j\rightarrow\nn&&
\frac{1}{N}{\sum}^\prime_{\v k,\v p, \v q\in \text{MBZ}}\sum_{a,b,c,d} \Big\{U_{\text{cd}}^{\text{ba}}(\v q)C_{\v k+\v q,a}^+ C_{\v k,b}C_{\v p-\v q,c}^+ C_{\v p,d}+ U_{\text{cd}}^{\text{ba}}(\v q+\v Q)C_{\v k+\v q+\v Q,a}^+ C_{\v k,b}C_{\v p-\v q-\v Q,c}^+ C_{\v p,d}\nn&&+ U_{\text{cd}}^{\text{ba}}(\v q)C_{\v k+\v q+\v Q,a}^+ C_{\v k+\v Q,b}C_{\v p-\v q,c}^+ C_{\v p,d}+ U_{\text{cd}}^{\text{ba}}(\v q+\v Q)C_{\v k+\v q,a}^+ C_{\v k+\v Q,b}C_{\v p-\v q-\v Q,c}^+ C_{\v p,d}\nn&& +U_{\text{cd}}^{\text{ba}}(\v q)C_{\v k+\v q,a}^+ C_{\v k,b}C_{\v p+\v Q-\v q,c}^+ C_{\v p+\v Q,d} + U_{\text{cd}}^{\text{ba}}(\v q+\v Q)C_{\v k+\v q+\v Q,a}^+ C_{\v k,b}C_{\v p-\v q,c}^+ C_{\v p+\v Q,d}\nn&&+U_{\text{cd}}^{\text{ba}}(\v q)C_{\v k+\v q+\v Q,a}^+ C_{\v k+\v Q,b}C_{\v p+\v Q-\v q,c}^+ C_{\v p+\v Q,d}+ U_{\text{cd}}^{\text{ba}}(\v q+\v Q)C_{\v k+\v q,a}^+ C_{k+\v Q,b}C_{\v p-\v q,c}^+ C_{\v p+\v Q,d}\Big\}\label{mint}
\ee
In \Eq{mint} $N$ is the total number of unit cells, $\v Q=(\pi,\pi)$ is the AF ordering wavevector, a,b,c,d are spin induces. The $U_{\text{cd}}^{\text{ba}}(\v q)$ in \Eq{mint} is given by
\be U_{\text{cd}}^{\text{ba}}(\v q)=\left(\vec{\sigma}_{\text{ab}}\cdot\vec{\s}_{\text{cd}}\right)J(\v q),~~J(\v q)=J(\cos q_x+\cos q_y),\ee where $\vec{\s}$ are the spin Pauli matrices.
The operator $\psi_{\v k}$ in \Eq{S1} is related to $C_{\v k\s}$ in \Eq{mint} by
\be
\psi_{\v k\s}=\phi_{\v k,\s,n(\v k),1}(\v k)C_{\v k,\s}+\phi_{\v k,\s,n(\v k),2}(\v k)C_{\v k+\v Q,\s}.\label{psi}\ee Here $\phi_{n,\s}^T=(\phi_{n,\s,1},\phi_{n,\s,2})$ us the n$^{\rm th}$ ($n=1,2)$ eigenvector of
\be
{\cal H}_\s(\v k)=\left(
          \begin{array}{ll}
            \e_0(\v k) & \s~m \\
            \s~m & \e_0(\v k+\v Q) \\
          \end{array}
        \right)\label{bm}\ee given in the main text.
In \Eq{psi} $n(\v k)$ in the subscript of $\phi$ denotes  is the index of the band that is closest to the fermi energy at momentum $\v k$. After some algebra \Eq{mint} can be reduced to
\be
V_{\text{eff}}=\frac{1}{N}{\sum}^\prime_{\v k,\v p, \v q\in \text{MBZ}}\sum_{a,b,c,d}\Big\{ 2J(\v q)K_{\text{ba}}^{\text{ba}}(\v k,\v p;\v q)\psi _{\v k+\v q,a}^+ \psi _{\v p-\v q,b}^+ \psi _{\v p,a} \psi _{\v k,b}-J(\v q)K_{\text{aa}}^{\text{bb}}(\v k,\v p;\v q)\psi _{\v k+\v q,b}^+ \psi _{\v p-\v q,a}^+ \psi _{\v p,a}\psi _{\v k,b}\Big\},\ee
where
\be
K_{\text{cd}}^{\text{ba}}(\v k,\v p;\v q)&&=\Big\{\phi _{\v k+\v q,a,1}^*\phi _{\v k,b,1} \phi _{\v p-\v q,c,1}^* \phi _{\v p,d,1}-\phi _{\v k+\v q,a,2}^*\phi _{\v k,b,1}\phi _{\v p-\v q,c,2}^* \phi _{\v p,d,1}+\phi _{\v k+\v q,a,2}^*\phi _{\v k,b,2} \phi _{\v p-\v q,c,1}^* \phi _{\v p,d,1}\nn&&-\phi _{\v k+\v q,a,1}^*\phi _{\v k,b,2} \phi _{\v p-\v q,c,2}^* \phi _{\v p,d,1} +\phi _{\v k+\v q,a,1}^* \phi _{\v k,b,1}\phi _{\v p-\v q,c,2}^* \phi _{\v p,d,2}-\phi _{\v k+\v q,a,2}^* \phi _{\v k,b,1}\phi _{\v p-\v q,c,1}^* \phi _{\v p,d,2}\nn&&+\phi _{\v k+\v q,a,2}^*\phi _{\v k,b,2} \phi _{\v p-\v q,c,2}^* \phi _{\v p,d,2}-\phi _{\v k+\v q,a,1}^* \phi _{\v k,b,2}\phi _{\v p-\v q,c,1}^* \phi _{\v p,d,2}\Big\}.
\label{bk}\ee
In \Eq{bk} we have omitted $n(\v k)$ in the subscript of $\phi$ because in the momentum thin shell $\v k$ fixes $n(\v k)$. %Moreover in arriving at \Eq{bk} we have used the fact that ${\cal H}_\s(-\v k)={\cal H}_\s(\v k)$ hence it is possible to choose the gauge of the wavefunction so that $\phi_{-\v k,\s}=\phi_{\v k,\s}$.

In the Cooper pairing channel the surviving term of \Eq{bk} are
\be V_{\text{eff}}\ra -{1\over N}{\sum}^\prime_{\v k,\v p, \v q\in \text{MBZ}}\sum_{a,b,c,d}\Lambda _{\text{ab}}^{\text{ab}}(\v k;\v p)\psi _{-\v k,b}^+\psi _{\v k,a}^+\psi _{\v p,a}\psi _{-\v p,b},\label{pv}\ee
 where
 \be
 \Lambda _{\text{ab}}^{\text{ab}}(\v k;\v p)=2J(\v p+\v k)K_{\text{ba}}^{\text{ba}}(-\v p,\v p;\v p+\v k)+J(\v p-\v k)K_{\text{aa}}^{\text{bb}}(-\v p,\v p;\v p-\v k).\ee

 Upon SC mean-field factorization \Eq{pv} becomes
\be V_{\text{eff}}\ra -{\sum}^\prime_{\v k,\v p\in {\rm MBZ}}\sum_{a,b,c,d} \Lambda _{\text{ab}}^{\text{ab}}(\v k;\v p)\left\{\psi _{-\v k,b}^+\psi _{\v k,a}^+\Delta _{\text{ab}}(\v p)+\Delta _{\text{ab}}^*(\v k)\psi _{\v p,a}\psi _{-\v p,b}-\Delta _{\text{ab}}^*(\v k)\Delta _{\text{ab}}(\v p)\right\}.\ee
where due to fermion anticommutation relation $ \Delta_{ab}(\v k)=-\Delta_{ba}(-\v k)$. Thus
\be
\Delta(\v k)=\left(
               \begin{array}{cc}
                 \Delta_{t,1}(\v k) & \Delta_s(\v k)+\Delta_{t,3}(\v k) \\
                 -\Delta_s(\v k)+\Delta_{t,3}(\v k) & \Delta_{t,2}(\v k) \\
               \end{array}
             \right).\label{dt}\ee
where $\Delta_s(\v k)$ is even in $\v k$ and $\Delta_{t,j}(\v k)$ are odd in $\v k$. In the literature \Eq{dt} is often written as
\be
\Delta(\v k)=[\psi(\v k)\s_0+\vec{d}(\v k)\cdot\vec{\s}](\imth\s_y),\ee where $\psi(\v k)=\Delta_s(\v k), d_x(\v k)=\frac{1}{2} \left[-\Delta _{t,1}(\v k)+\Delta _{t,2}(\v k)\right], d_y(\v k)=\frac{1}{2i}\left[\Delta _{t,1}(\v k)+\Delta _{t,2}(\v k)\right], d_z(\v k)=\Delta _{t,3}(\v k)$.

We then ``integrate out'' the electrons and keep up to the quadratic terms in $\Delta$'s.  The result is the following free energy form
\be
F={1\over N}{\sum}'_{\v k,\v p\in {\rm MBZ}}\sum_{a,b}\Delta_{ab}^*(\v k)K_{ab}(\v k,\v p;T)\Delta_{ab}(\v p),\ee where
\be K_{ab}(\v k,\v p;T)=\Lambda _{\text{ab}}^{\text{ab}}(\v k;\v p)-\frac{2}{N}{\sum}^\prime_{\v q\in{\rm MBZ}}\Lambda _{\text{ab}}^{\text{ab}}(\v k;\v q)\chi(\v q;T)\Lambda _{\text{ab}}^{\text{ab}}(\v q;\v p)\ee
where the temperature ($T$)-dependent free fermion pair susceptibility is given by
\be
\chi(\v k;T)%=\int d\tau_0^\beta\Big{\langle}T_\tau\left(\sum_{s_1,s_2}\e_{s_1s_2}\psi^\dagger_{-\v q,s_1}\psi^\dagger_{\v q,s_2}\right)_\tau\left(\sum_{s_3,s_4}\e_{s_3 s_4}\psi_{\v q,s_4} \psi_{-\v q,s_3}\right)_0\Big{\rangle}_{0,c}
\propto{1-2f(\e(\v k))\over \e(\v k)}.\ee Here the proportionality constant is un-important for our purposes as it can be absorbed into $J$ (see below).

The leading (sub-leading) gap functions are the eigenfucntions of $\chi(\v q;T)\Lambda _{\text{ab}}^{\text{ab}}(\v q;\v p)$ with the largest (second largest) eigenvalue among all $a,b$ (The proportionality constant in $\chi_T$ changes all eigenvalues by the same multiplicative constant but not the eigenfunctions.) These are the order parameters which will first (second) become unstable as $J$ increases (at a temperature $T$ much less than the thickness of the energy shell). These eigenfunctions are obtained numerically after discretizing the momentum space enclosed by the energy shell (under such discretization $\chi(\v q;T)\Lambda _{\text{ab}}^{\text{ab}}(\v q;\v p)$ becomes a matrix for each pair of $(a,b)$). We diagonalize each matrix then average the eigenfunctions along the direction perpendicular to the fermi surface. This leads to the results presented in the text.

%\section{Aspects of topological defects}

%\bibliographystyle{apsrev}
%%\bibliography{F://Research//Ctex//Bib//mybib//bibs}
%\bibliography{E://Dropbox//notes//bibs}